\documentclass[final,3p,times,twocolumn]{elsarticle}

\usepackage{graphicx}

\usepackage{amssymb}

\usepackage{lineno}

\usepackage{chngpage}

\biboptions{numbers,square,comma,sort&compress}

\usepackage{hyperref}

\usepackage{hypernat}

\journal{Physics Letters B}

\begin{document}

\begin{frontmatter}


\title{\vspace{-0.8cm} \flushright{\normalsize{CERN-PH-EP-2011-129}} \\ \center{Search for the Standard Model Higgs boson in the two photon decay channel with the ATLAS detector at the LHC}}

\author{The ATLAS Collaboration}

\address{}

\begin{abstract}

A search for the Standard Model Higgs boson in the two photon decay channel is reported, using 1.08~fb$^{-1}$ of 
proton-proton collision data at a centre-of-mass energy of 7~TeV recorded by the ATLAS detector.
No significant excess is observed in the investigated mass range of 110-150~GeV.
Upper limits on the cross-section times branching ratio of between 2.0 and 5.8 times the Standard Model prediction
are derived for this mass range.

\end{abstract}

\end{frontmatter}


\section{Introduction}

The search for the Standard Model Higgs boson~\cite{higgs1,higgs2,higgs3} is one of the key goals of the
Large Hadron Collider (LHC) at CERN. 
The allowed Higgs boson mass ($m_H$) is constrained at the 95\% confidence level by a lower limit 
of 114.4~GeV from the
LEP experiments~\cite{lep} and an excluded region between 156 and 177 GeV from the Tevatron
experiments~\cite{tevatron1,tevatron2}.
First results have been reported by the ATLAS experiment in a variety of channels~\cite{atlas2010}
and the CMS experiment~\cite{cms2010} using the data recorded in 2010, which correspond to an
integrated luminosity about thirty times smaller than the 2011 dataset used in this analysis.

In the low mass range, from the LEP limit to $m_H$ $\approx$~140~GeV, one of the most promising search channels
at the LHC is the rare decay of the Higgs boson into a pair of photons. Despite the low branching ratio ($\approx$~0.2\%),
this channel provides good experimental sensitivity in the mass region below 150 GeV.
The results presented in this paper are based on proton-proton collision data taken
at $\sqrt{s}$~=~7~TeV by the ATLAS experiment between April and June 2011.

The data analysis proceeds by selecting photon pairs with tight identification and isolation cuts to minimize backgrounds
other than direct diphoton production. 
A narrow peak in the reconstructed invariant mass
distribution is searched for over a large, smooth background whose
normalisation and shape are left free in a maximum likelihood fit.
To increase the sensitivity, the sample is divided into five categories based on the presence of photon conversions and
on the photon impact point on the calorimeter, with different invariant mass resolutions
and signal-to-background ratios for the different categories.

The results of the fit are compared to the prediction from the Standard Model using the Higgs boson production cross-section
and branching ratio from Ref.~\cite{higgs-theory}. Limits on the production cross-section relative to the Standard Model value
are then derived as a function of the hypothesised Higgs boson mass.
Although with the current dataset the analysis is not yet sensitive 
to the predicted rate for a Standard Model Higgs boson, the
limits on the yield in this decay channel improve on those obtained 
in the same channel by the Tevatron experiments~\cite{tev-gg1,tev-gg2,tev-gg3},
and are sensitive to possible enhancements in the Higgs boson production and decay rate compared with the Standard Model expectations.

\section{Experimental setup and data set}

The ATLAS detector is described in detail in Ref.~\cite{ATLAS_detector}.
The main subdetectors relevant to this analysis 
are the calorimeter, in particular its
electromagnetic section, and the inner tracking detector.

The electromagnetic calorimeter is a
lead-liquid argon sampling calorimeter with accordion geometry. 
It is divided into a barrel section covering the pseudorapidity\footnote{
ATLAS uses a right-handed coordinate system with its origin at 
the nominal interaction point (IP) in the centre of the detector 
and the $z$-axis along the beam pipe. 
The $x$-axis points from the IP to the centre of the LHC ring, 
and the $y$ axis points upward.
 Cylindrical coordinates $(r,\phi)$ are used in the transverse plane, 
$\phi$ being the azimuthal angle around the beam pipe. 
The pseudorapidity is defined in terms of the polar angle 
$\theta$ as $\eta=-\ln\tan(\theta/2)$.} 
region $|\eta|< 1.45$ and two end-cap sections covering the pseudorapidity
region $1.375<|\eta|<3.2$. 
It has three longitudinal layers.
The first one, with a thickness between 3 and 5 radiation lengths, has
a high 
granularity in $\eta$ (between 0.003 and 0.006 depending
on $\eta$, with the exception of the regions $1.4<|\eta|<1.5$ and
$|\eta|>2.4$), sufficient to provide 
discrimination between single photon showers and two photons
from a $\pi^0$ decay.
The second layer
has a thickness of around 17 radiation lengths 
and a granularity of $0.025\times0.025$ in $\eta\times\phi$.
A third layer, with a thickness varying between 4 and 15 radiation lengths, is
used to correct for leakage beyond the calorimeter for high energy showers.
In front of the calorimeter, a thin presampler layer, covering
$|\eta|<1.8$, is used to correct for
fluctuations in upstream energy losses.
The sampling term $a$ of the energy resolution, $\sigma(E)/E$
$\approx$ $a/\sqrt{E~{\rm (GeV)}}$, varies between 9\% and 14\% as a
function of $|\eta|$ for unconverted photons~\cite{ATLAS_EMC_calibration}.
It reaches up to 20\% for converted photons near $|\eta|$ of 1.3 where the upstream
material effect is the largest.
The sampling term is the largest contribution to the resolution
up to about 100 GeV, where the constant term starts to dominate.
After 0.17~fb$^{-1}$ of data were accumulated, 
some calorimeter cells could not be read out.
The affected region size is
$\Delta\eta\times\Delta\phi$$\approx$1.5$\times$0.2 in the barrel electromagnetic calorimeter, resulting
in an acceptance loss for diphoton candidates of about~3\%.
A hadronic sampling calorimeter is located behind the
electromagnetic calorimeter. 
It is made of steel and scintillating tiles in the barrel section,
and of copper and liquid argon in the end-cap.

The inner detector consists of three subsystems: 
at small radial distance $R$ from the beam axis ($5<R<15$ cm), pixel
silicon detectors are arranged in three cylindrical layers in 
the barrel and in three disks in each end-cap; at intermediate radii
($30<R<56$ cm), double layers of single-sided silicon microstrip
detectors are used, organised in four cylindrical layers in the barrel
and nine disks in each end-cap; at larger radii ($56<R<107$ cm), a
straw tracker with transition radiation 
capabilities is used.
These three systems are immersed in a 2 T axial magnetic field.
The silicon pixel and 
strip subsystems cover the range $|\eta|<2.5$, while the
transition radiation tracker acceptance is limited to the range $|\eta|<2.0$. 
The inner detector allows reconstruction of secondary
vertices, in particular of photon conversions occurring
in the inner detector material up to a radius of $\approx$~80~cm.

The total amount of material in front of the first active layer of the
electromagnetic calorimeter (including that in the presampler) varies between
2.5 and 6 radiation lengths as a function of pseudorapidity, excluding 
the transition region ($1.37<|\eta|<1.52$) between the barrel and the
end-caps.

Data used in this analysis were selected using a di-photon trigger with
a 20~GeV transverse energy threshold on each photon. At the first trigger level, which uses reduced granularity,
two clusters with transverse energies above 14~GeV
are required in the electromagnetic calorimeter.
At the higher trigger levels, loose photon identification cuts are applied using the
full calorimeter granularity. This trigger has an efficiency greater than 99\% for the signal after the
final event selection.

In these data, the instantaneous luminosity varies between $\approx$10$^{32}$cm$^{-2}$s$^{-1}$
and  $\approx$10$^{33}$cm$^{-2}$s$^{-1}$ with a bunch spacing of 50~ns.
The average number of collisions per bunch
crossing is around~6. 
Collisions in the same bunch crossing as the signal (in-time pileup) or in other bunch crossings within the detector sensitive time (out-of-time pileup) influence the event reconstruction.
The inner detector is only sensitive to in-time pileup
while the electromagnetic calorimeter is sensitive to pileup within a $\approx$~450~ns time window.

The application of beam, detector, and data-quality requirements to the recorded data results in a data
sample corresponding to a total integrated
luminosity of (1.08$\pm$0.04)~fb$^{-1}$~\cite{lumi2011}.

\section{Simulated samples}
\label{sec:theory}

The Higgs boson signal from the dominant gluon fusion production process (corresponding to 86\% of the production cross-section
for a Higgs boson with a mass of 120~GeV) is generated with POWHEG~\cite{powheg_gg}.
MC@NLO~\cite{mcnlo} is used as a cross-check. POWHEG~\cite{powheg_vbf} is also used to generate
the signal events from the sub-leading vector boson fusion process (7\% of the 
cross-section at 120~GeV).
For the other production modes, namely associated production with a $W$ or $Z$ boson or a 
$t\bar{t}$ pair, PYTHIA~\cite{pythia} is used.

The predicted signal is normalised using NNLO cross-sections
for the gluon fusion process~\cite{Harlander:2002wh,Anastasiou:2002yz,Ravindran:2003um,
Anastasiou:2008tj,deFlorian:2009hc}, the vector boson fusion process~\cite{Bolzoni:2010xr}, the
associated production with a $W$ or $Z$ boson~\cite{Brein:2003wg} and NLO cross-section for the associated
production with a $t\bar{t}$ pair~\cite{Dawson:2002tg}.
The NLO electroweak corrections are applied to the gluon fusion~\cite{Aglietti:2004nj,Actis:2008ug}, 
vector boson fusion~\cite{Ciccolini:2007jr,Ciccolini:2007ec},
and the associated production with a $W$ or $Z$ boson~\cite{Ciccolini:2003jy} processes.
The uncertainty on the theoretical cross-section is estimated~\cite{higgs-theory}
to be $^{+20}_{-15}$\%, mostly due to the renormalisation and factorisation scale variations and
the uncertainties in the parton distribution functions~\cite{Botje:2011sn,Lai:2010vv,Martin:2009iq,Ball:2011mu}.
The Higgs boson decay branching fractions are taken from Refs.~\cite{higgs-theory,Djouadi:1997yw}. The uncertainty
on the branching ratio to two photons is negligible compared with the cross-section uncertainty.

Signal events are generated in steps of 5 GeV for Higgs boson masses in the range of 110-150~GeV.
PYTHIA and  ALPGEN~\cite{alpgen} have been chosen
to generate the background samples,  which are, however, only used for cross-checks and
not to extract the final results.

All Monte Carlo (MC) samples are processed through a complete simulation
of the ATLAS detector~\cite{atlasSim} using the GEANT4 programme~\cite{geant4}.
Pileup effects are simulated by overlaying each MC event with a variable number of
MC inelastic $pp$ collisions, taking into account both 
in-time and out-of-time pileup
and the LHC bunch train structure.
MC events are weighted to have the same distribution of average number of interactions
per bunch crossing as in the data.

\section{Photon reconstruction, event selection and backgrounds}

\subsection{Photon reconstruction}

Photon reconstruction is seeded by energy clusters
in the electromagnetic calorimeter with transverse energies exceeding 2.5~GeV 
in projective towers of size 0.075$\times$0.125 in $\eta\times\phi$ made from the presampler
and the three electromagnetic calorimeter layers.
These energy clusters are then matched to tracks that are reconstructed in the
inner detector and extrapolated to the calorimeter.
Clusters without matching tracks are classified as unconverted
photon candidates.
Clusters matched to either pairs of tracks which are consistent with the 
hypothesis of a photon conversion or single tracks without hits 
in the pixel layer nearest to the beam pipe are considered as converted photon candidates.
The photon reconstruction efficiency is $\approx$98\%. 

The energy measurement is made in the electromagnetic calorimeter
using a cluster size which
depends on the photon classification.
In the barrel, a size of 0.075$\times$0.125 in $\eta\times\phi$
is used for unconverted photons and 0.075$\times$0.175 for converted photon
candidates, to account for the larger spread of the shower in $\phi$ for converted photons
due to the magnetic field.
In the end-cap, a cluster size of 0.125$\times$0.125 is used for all candidates. 
A dedicated energy calibration~\cite{ATLAS_EMC_calibration}
is applied to account for upstream energy losses, lateral leakage
and longitudinal leakage, separately for converted and unconverted photon candidates.

The final energy calibration is determined from $Z\rightarrow e e$ decays,
resulting in  $\eta$-dependent correction factors of the order of $\pm$1\%.
After this calibration procedure, the constant term in the energy resolution is
estimated to be 1.1$^{+0.5}_{-0.6}$\% in the barrel region and 1.8$^{+0.5}_{-0.6}$\% in the 
end-cap region~\cite{egamma2010}. The energy resolution in the simulation is adjusted to
match these values.

Photon identification is based on the lateral
and longitudinal energy profiles of the shower in the calorimeter~\cite{photon2010}.
The photon candidate is required to deposit only a small fraction of its energy in the hadronic calorimeter.
The transverse shower shape in the second layer of the electromagnetic calorimeter
needs to be consistent with that expected for a single electromagnetic shower. Finally, the high granularity
first layer is used to discriminate single photons from overlapping photon pairs
from neutral meson decays produced in jet fragmentation, which are the main background source.
Based on these criteria, a set of tight identification cuts, different for converted and unconverted
candidates, is applied.

To take into account small differences in shower shapes between data and simulation, the shape
variables are shifted in the simulation before the identification cuts are applied.
The photon identification efficiency ranges typically from 75\% to 90\% for transverse energies
between 25 and 100~GeV.

To increase the background rejection, an isolation cut is applied. The isolation variable~\cite{photon2010}
is computed by summing the transverse energy in calorimeter cells in a cone of radius 0.4
in the  $\eta\times\phi$ space around the photon candidate. 
Cells in the electromagnetic calorimeter within 0.125$\times$0.175 from the shower barycentre are excluded from
the sum. The small photon energy leakage outside the excluded cells is evaluated as a function of the transverse
energy in simulated samples and is subtracted from the isolation variable. To reduce the effect from the underlying
event and pileup, the isolation is further corrected using a method suggested in Ref.~\cite{Cacciari:area}:
for each of the two different pseudorapidity
regions $|\eta|<1.5$ and $1.5<|\eta|<3.0$, low energy jets are used
to compute an ``ambient'' energy density, which is then multiplied by the
area of the isolation cone and subtracted from the isolation energy.

In the following, photon candidates having isolation transverse energies lower
than 5 GeV are considered as isolated. 
The isolation cut efficiency is checked in data using a control sample of $Z\rightarrow e e$ events.
The per-event efficiency of requiring both electrons to be isolated is found to be 3\% lower in the
data than in the simulated samples. In the MC, the isolation cut efficiency is found to be the
same for $Z\rightarrow e e$ and $H \rightarrow \gamma \gamma$ events ($\approx$93\%). 
The number of events predicted by the simulation
after the isolation cut is therefore reduced by 3\%.

\subsection{Event selection}

Two photon candidates are required to pass tight identification criteria, to be isolated,
and to be within the region
$|\eta|<$2.37, excluding 1.37$<|\eta|<$1.52, where the first calorimeter layer has high granularity.
The highest and second highest photon transverse energies are required be
above 40 and 25~GeV respectively.
Both photons must be clear of
problematic regions in the calorimeter.
As the goal is to investigate Higgs boson mass hypotheses between 110 and 150~GeV,
the invariant mass of the photon pair is required to be within 100-160~GeV.
After these cuts 5063 events remain in the selected data sample.

The acceptance of the kinematic cuts, as estimated with generated photons in the MC signal samples,
is 60\% for the dominant gluon fusion process for a mass of 120~GeV.   
The overall event selection efficiency, taking into account both kinematic cut acceptance and reconstruction
and identification efficiencies, is 39\%.
The event selection efficiency is slightly larger in the vector boson fusion process.
It is somewhat smaller in the associated production mode.
It increases
with the Higgs boson mass from 34\% at 110~GeV to 43\% at 150~GeV.

To enhance the sensitivity of the analysis, the data sample is split in five categories, with different
invariant mass resolutions and different signal-to-background ratios:
\begin{itemize}
\item Unconverted central (8\% of the candidates): Both photons are unconverted and in the central part of the barrel calorimeter ($|\eta|<$0.75).
This is the category with the best invariant mass resolution and the best signal-to-background ratio;
\item Unconverted rest (28\% of the candidates): Both photons are unconverted and at least one photon does not lie in the central part of the barrel calorimeter;
\item Converted central (7\% of the candidates): At least one photon is converted and both photons are in the central part of the barrel calorimeter;
\item Converted transition (16\% of the candidates): At least one photon is converted and at least one photon is near the transition between barrel and end-cap
calorimeter (1.3$<|\eta|<$1.75). Given the larger amount of material in this region, the energy resolution, in particular for
converted photons, can be significantly degraded;
\item Converted rest (41\% of the candidates): All other events with at least one converted photon.
\end{itemize}

Table~\ref{table:yield} shows the cross-section times branching ratio to two photons, the expected total and per category signal yields for
1.08~fb$^{-1}$ for different Standard Model Higgs boson mass hypotheses.
Using these categories improves the signal sensitivity of the analysis by around 15\% for a 120~GeV Higgs boson mass
compared with a fully inclusive analysis.

\begin{table}[h!]
\begin{footnotesize}
 \begin{center}
   \hspace{1.0cm}\caption{Cross-section times branching ratio  and expected numbers of signal
events after all cuts (total and per category), for various Higgs boson masses 
and for an integrated luminosity of 1.08~fb$^{-1}$.}
 \begin{tabular}{lccccc}
   \hline\hline
   $m_H$ [GeV]       & 110   & 120   & 130   & 140   & 150  \\
   \hline
   $\sigma\times$BR [fb] &  45  & 43  & 37  & 27   & 16 \\
   \hline
   Signal yield          & 17.0 & 17.6 & 15.8 & 12.1 & 7.7  \\
    \hline
   Unconverted central & 2.6 & 2.6 & 2.3 & 1.7 & 1.1 \\
   Unconverted rest & 4.6 & 4.7 & 4.2  & 3.4  & 2.1 \\
   Converted central & 2.0 & 2.0  & 1.7 & 1.3 & 0.8 \\
   Converted transition & 2.3 & 2.2 & 2.1  & 1.5  & 1.0  \\
   Converted rest       & 5.6 & 6.0 & 5.6  & 4.2 & 2.7  \\
   \hline\hline
 \end{tabular}
  \label{table:yield}
\end{center}
\end{footnotesize}
\end{table}

\subsection{Invariant mass reconstruction}

In addition to the energies, the angle between the photons
is needed for the computation of the diphoton invariant mass.
This angle is determined from the interaction vertex position and the 
photon impact points in the calorimeter. The resolution of
the angle measurement is dominated by the reconstruction of 
the primary vertex $z$ position.
The RMS vertex spread in the $z$ direction is $\approx$5.5 cm, and 
a more accurate event-by-event estimate is performed to reduce
the impact on the invariant mass resolution. Given the non-negligible level
of pileup in the 2011 data, the determination of the vertex position is 
based only on the photon candidates, without relying on other charged 
tracks in the event.
For converted photons with tracks
having a precise measurement in the $z$ direction, the vertex position is estimated from the intercept
of the line joining the reconstructed conversion position and the calorimeter impact point with the beam line. For all other
photons, the vertex position is estimated from the shower position measurements in the first and second layers
of the electromagnetic calorimeter, which can be used to calculate the photon direction. Finally, the vertex positions from both photons
are combined taking also into account the average beam spot position in $z$. When both photons are unconverted, the
typical vertex position resolution is $\approx$ 1.6~cm in $z$. 
The resolution is better in events with converted photons. The resulting impact
of the angle measurement on the invariant mass resolution is negligible compared
to the contribution from the photon energy resolution.

\begin{figure}[h!]
\begin{center}
\includegraphics[width=0.45\textwidth]{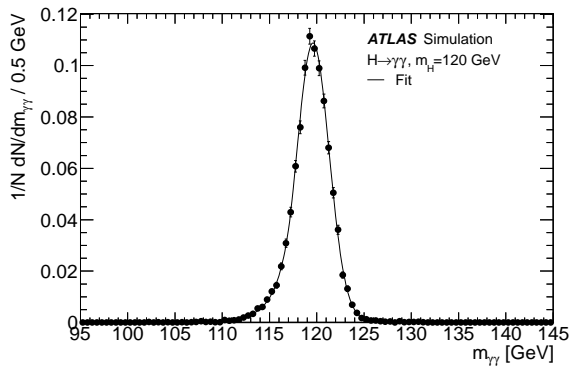}
\end{center} 
\caption{Distribution of the reconstructed diphoton invariant mass of
  a simulated 120~GeV mass Higgs boson signal, for all categories together. The line shows
the fit of the mass resolution using the function described in the text. The core component of the mass
 resolution is 1.7~GeV}
\label{fig:massResolution}
\end{figure}

Figure~\ref{fig:massResolution} shows the invariant mass distribution for simulated Higgs boson events with mass 120~GeV.
The mass resolution for the signal is modelled by the sum of a Crystal Ball function~\cite{crystalBall} (for the bulk of the
events which have a narrow Gaussian spectrum in the peak region and tails toward lower reconstructed mass) 
and a Gaussian distribution with a wide sigma (to model the far outliers in the distribution). 
The Crystal Ball function is defined as:

\begin{footnotesize}
\begin{center}
\begin{displaymath}
 N\cdot\left\{ 
    \begin{array}{ll}
      e^{-t^{2}/2} & \mbox{if $t>-\alpha_{CB}$}  \\ 
      (\frac{n_{CB}}{\alpha_{CB}})^{n_{CB}} \cdot e^{-\alpha_{CB}^2/2} \cdot (\frac{n_{CB}}{\alpha_{CB}}-\alpha_{CB}-t)^{-n_{CB}}  & \mbox{otherwise} 
    \end{array} 
  \right.  
\end{displaymath}
\end{center}
\end{footnotesize}
where $t = (m_{\gamma\gamma}-\mu_{CB})/\sigma_{CB}$, $N$ is a normalisation parameter, $\mu_{CB}$ is the peak of the narrow Gaussian distribution, 
$\sigma_{CB}$ represents
the Gaussian resolution for the core component, and $n_{CB}$ and $\alpha_{CB}$ parametrise the non-Gaussian tail.\\

The core component of the mass resolution, $\sigma_{CB}$, ranges
from 1.4~GeV in the ``Unconverted central'' category to 2.1~GeV in the ``Converted transition'' category. 
The non-Gaussian contributions to the mass resolution arise mostly from converted photons 
with at least one electron losing a significant fraction of its energy through bremsstrahlung in the inner detector material.

\subsection{Sample composition}

 The main background components are the diphoton production,
the photon-jet production with one fake photon from jets fragmenting
into a high energy  $\pi^0$, the dijet production with two fake photons,
and Drell-Yan events where
both electrons are misidentified as photons.
A measurement of the diphoton production cross-section with 2010 ATLAS data can
be found in Ref.~\cite{diphoton2010}, where the techniques used to estimate the purity
of the sample are described in more detail.
Although the final result does not rely on it, a quantitative understanding
of the sample composition is an important cross-check of the diphoton selection
procedure.

 A method based on the use of control regions for two discriminating variables
is applied to measure the contributions of fake photon background directly
from the data.  This method exploits relaxed isolation and photon identification
cuts to estimate the fake components, by relying on the fact that the rejections from these
two cuts are almost independent. It is a generalisation of the
method used in Ref.~\cite{photon2010}.
The Drell-Yan background is estimated by measuring the probability for an electron to be reconstructed as a photon candidate
with $Z$ events and applying this probability to the observed yield of Drell-Yan events at high mass.

The number of diphoton events in the 100-160~GeV mass range is found to be 3650$\pm$100$\pm$290,
where the first uncertainty is statistical and the second is systematic.
The systematic uncertainty arises from the definition of the relaxed identification control region,
the possible correlations between isolation and identification variables, and the fraction
of real photons leaking into the background control regions. 
The extracted yields of photon-jet and dijet are 1110$\pm$60$\pm$270 and 220$\pm$20$\pm$130 events respectively.
The Drell-Yan background, which is most prominent in the categories with at least one converted photon,
is estimated to be 86$\pm$1$\pm$14 events in the mass range of 100-160~GeV.

\begin{figure}[h!]
\begin{center}
\includegraphics[width=0.45\textwidth]{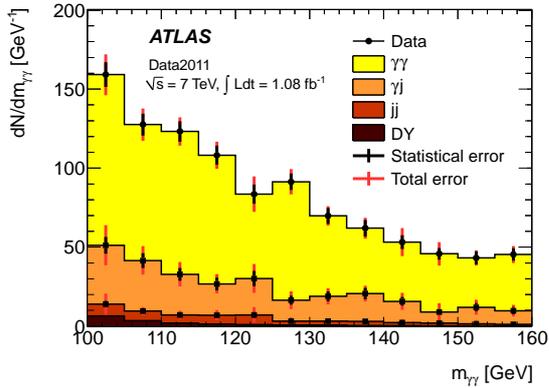}
\end{center} 
\caption{Diphoton, photon-jet, dijet and Drell-Yan contributions to the diphoton candidate invariant
mass distribution, as obtained from a data-driven method. The various components are stacked on top of each other. The error bars correspond
to the uncertainties on each component separately.}
\label{fig:massComposition}
\end{figure}

Figure~\ref{fig:massComposition} shows the extracted components of the diphoton, photon-jet, dijet and Drell-Yan
processes.
The purity of the sample (fraction of diphoton events) is about 72\%.
The measurement of the purity has also been made separately in each category, and
ranges from 69\% to 83\%.

Other methods have been used to cross-check the purity estimate, in particular
using template fits of the photon isolation distribution, where both signal and background
templates are derived from data. The results are in agreement with the results quoted here.

\section{Systematic uncertainties}
\label{sec:exp_syste}

Experimental systematic uncertainties affecting the extraction of the signal from the
diphoton invariant mass distribution related to the modelling of the signal 
can be classified in two types: uncertainties affecting the predicted yield and uncertainties
affecting the modelling of the mass resolution.

The uncertainties on the event yield are the following:
\begin{itemize}
\item The uncertainty from the photon reconstruction and identification efficiency amounts
to $\pm$11\% per event. It is estimated from data and MC differences in shower
shape variables, the impact of additional material in front of the calorimeter and the impact
of pileup on the photon shower shape variables.
\item The uncertainty on the isolation cut efficiency is taken as the difference
between data and MC found in $Z\rightarrow e e$ decays and amounts to $\pm$3\% per event.
\item The uncertainty on the photon trigger efficiency is $\pm$1\%. It comes from the
uncertainty in the measurement of the trigger efficiency for diphoton candidates using
control triggers and from possible differences between the trigger efficiency for photons from
Higgs boson decays and all diphoton candidates.
\item The uncertainty on the kinematic cut acceptance from the modelling of the Higgs boson transverse momentum
distribution is investigated with HQT~\cite{hqt} and RESBOS~\cite{resbos1,resbos2},
which account for all-orders soft-gluon resummation up to NNLL accuracy. The resulting
uncertainty is found to be at the level of $\pm$1\%.
\item The luminosity uncertainty is 3.7\%~\cite{lumi2011}. 
\end{itemize}
The total uncertainty on the expected signal event yield is $\pm$12\%.

The uncertainties on the invariant mass resolution are the following:
\begin{itemize}
\item The uncertainty on the cluster energy resolution comes from the uncertainty
on the sampling term, estimated to be 10\%, and from the uncertainty on the constant term,
which is estimated using $Z\rightarrow ee$ decays. Both uncertainties are taken into account
with their proper correlation from the Z control sample constraint. The uncertainty
on the cluster energy resolution amounts to a  $\pm$12\% relative uncertainty on the diphoton
invariant mass resolution.
\item The uncertainty on the photon energy calibration arising from the extrapolation of the electron energy
scale calibration is estimated from MC studies. 
The difference between the photon and the electron response in the calorimeter
comes from the material in front of the active part of the calorimeter. The uncertainty 
is estimated using simulations with a different amount of material in front of the
calorimeter and is found to be $\pm$6\% on the mass resolution.
\item The contribution of pileup fluctuations to the cluster energy measurement is checked using
random clusters in randomly triggered bunch crossings, with a frequency corresponding to expectations from the instantaneous luminosity.
The relative uncertainty 
on the mass resolution is found to be less than 3\%.
\item The uncertainty on the resolution of the photon angle measurement is studied with $Z\rightarrow ee$ decays.
The calorimeter-based direction measurement is compared with the much more precise track-based
direction measurement. In the barrel calorimeter, the resolution measured in data agrees
well with the one predicted by the simulation. In the end-cap region, the resolution measured in
data is $\approx$20\% worse than in the simulation. The impact of this difference 
is a  1\% relative uncertainty on the diphoton mass resolution.
\end{itemize}
The total relative uncertainty on the diphoton invariant mass resolution is thus $\pm$14\%.
This systematic uncertainty is applied to both the Crystal Ball and the wide Gaussian resolution
parameters.

These systematic uncertainties are taken as fully correlated between the different categories.
The impact of uncorrelated systematic uncertainties in the different categories and migration
between categories has been investigated
and found to be negligible.

The background is modelled by an exponentially falling invariant mass distribution.
Systematic uncertainties arise from possible deviations of the
background distribution from this assumed shape.
This has been estimated by checking how accurately the chosen
model fits different predicted diphoton mass distributions~\cite{diphox,resbos3} and comparing
different functional forms for the background model. 
The resulting uncertainty is
between $\pm$5 events at 110~GeV and $\pm3$ events at 150~GeV
for a Higgs boson mass signal region about 4~GeV wide.

\section{Results}

The data are compared to background and signal-plus-background hypotheses using 
a profile likelihood test statistic as described in Refs.~\cite{atlas2010,qmu}.
The exponentially falling invariant mass distribution used for the background model is determined
by two nuisance parameters per category (the normalisation and the exponential negative slope), which
are left free in the unbinned fit.
The signal is modelled with the mass resolution functions described above, one per category, fixing the fraction of events
in each category to the MC predictions. 
Table~\ref{table:table2} summarises the measured fractions of background events in each category, the predicted fractions of signal events and
the predicted core gaussian signal mass resolutions for a Higgs boson mass hypothesis of 120~GeV.

\begin{table}[h!]
\begin{footnotesize}
 \begin{center}
   \hspace{1.0cm}\caption{Fractions of background ($f_b$), predicted signal ($f_s$) and core gaussian mass resolution ($\sigma$) in each
category for a Higgs boson mass hypothesis of 120~GeV. The fractions are normalised to the total yield summed over categories.}
 \begin{tabular}{lccc}
   \hline\hline
category            &    $f_b$ & $f_s$ &  $\sigma$(GeV) \\
 \hline
Unconverted central &      7\%  &  15\%  &  1.4 \\
Unconverted rest    &     29\%  &  27\%  &  1.6 \\
Converted central   &      8\%  &  11\%  &  1.5 \\
Converted transition &    16\%  &  13\%  &  2.1 \\
Converted Rest      &     40\%  &  34\%  &  1.8 \\
  \hline\hline
 \end{tabular}
  \label{table:table2}
\end{center}
\end{footnotesize}
\end{table}

The fitted parameters for the signal are thus the overall signal strength relative
to the Standard Model prediction and the nuisance parameters on the predicted event yield and mass
resolution which have Gaussian constraints in the fit. The uncertainty on the predicted event yield includes both
the experimental systematic uncertainties described in section~\ref{sec:exp_syste} and the uncertainty on the predicted cross-section
described in section~\ref{sec:theory}.
The systematic uncertainty on the background shape
is included as another nuisance parameter with a Gaussian constraint in the fit.
From this fit, the best estimate of the signal yield is extracted, as well as the likelihood ratio (profile likelihood) between
any assumed signal yield (leaving the nuisance parameters free to maximise the likelihood) and the best estimate.
The fit is performed in 1~GeV steps
for the Higgs boson mass hypothesis, which is significantly smaller than the invariant mass resolution. The
signal parameters for these fine mass steps are interpolated from the fully simulated samples.

\begin{figure}[h!]
\begin{center}
\includegraphics[width=0.45\textwidth]{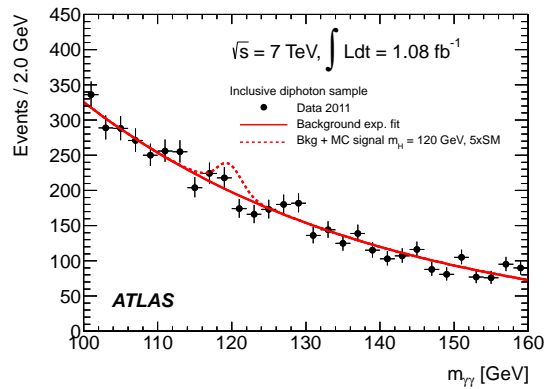}
\end{center} 
\caption{Distribution of the reconstructed diphoton mass. All five diphoton categories have been combined.
The exponential fit to the full sample of the background-only hypothesis, as well
as the expected signal for a Higgs boson mass of 120~GeV with five times the Standard Model predicted yield, are also shown for illustration.}
\label{fig:massExp}
\end{figure}

\begin{figure}[h!]
\begin{center}
\includegraphics[width=0.45\textwidth]{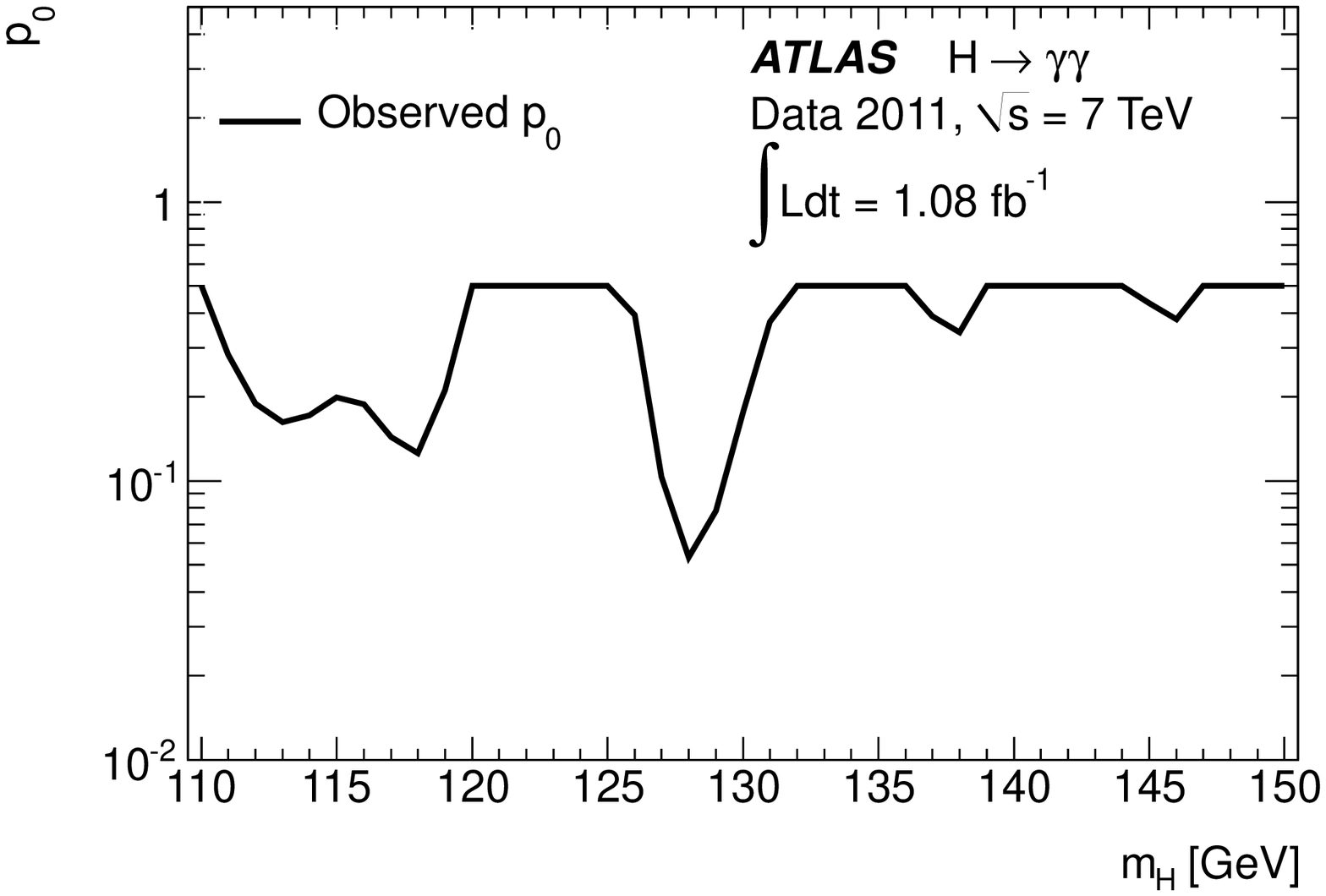}  \\
\vspace{-0.5cm}\hspace{1.1cm} \mbox{(a)} \\
\includegraphics[width=0.45\textwidth]{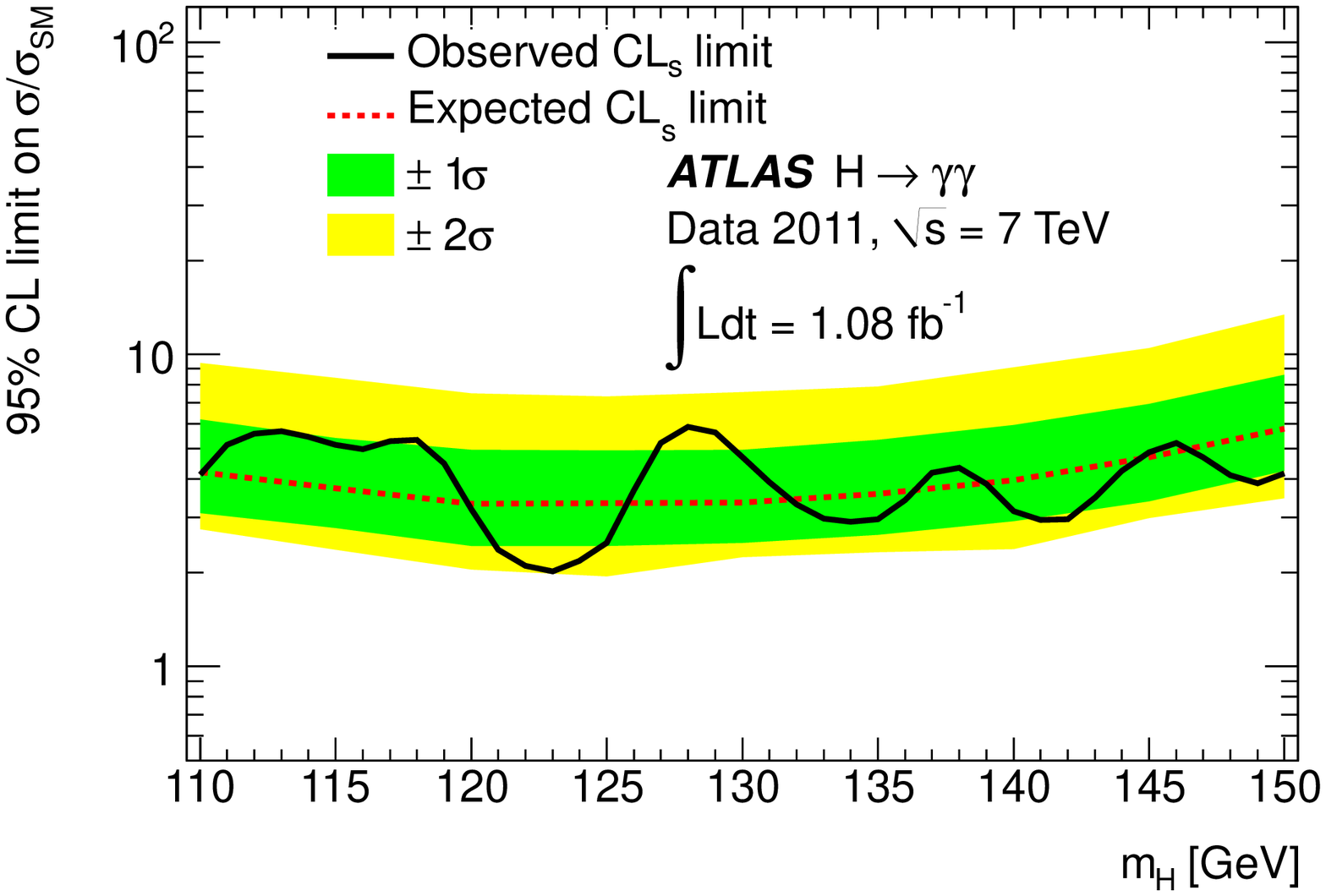} \\
\vspace{-0.5cm}\hspace{1.1cm} \mbox{(b)}
\vspace{-0.5cm}
\end{center}
\caption{ (a) $p$-value ($p_0$)  of the background-only hypothesis as a function of the investigated Higgs boson mass
(b)  95\% confidence level upper limits on a Standard Model Higgs boson production cross-section, relative to the
Standard Model cross-section,
as a function of Higgs boson mass hypothesis. The solid line shows the observed limit.
The dashed line shows the median expected limit for background-only pseudo-experiments, with the bands indicating the
expected fluctuations around this limit at the 1 $\sigma$ and 2 $\sigma$ levels.}
\label{fig:Results}
\end{figure}

Figure~\ref{fig:massExp} shows the reconstructed diphoton mass spectrum. No excess is visible.
This is quantified by the $p$-value of the background-only hypothesis, which gives the fraction of
background-only experiments that would have a profile likelihood ratio of the zero signal hypothesis relative
to the best-fitted signal strength at least as low as
the one found in the data. 
Negative signals are not allowed in the fit, so $p$-values above 0.5 are
truncated.
This $p$-value is shown in Figure~\ref{fig:Results}(a) as a function
of the hypothesised Higgs boson mass.
The minimal $p$-value, corresponding to the largest background upward fluctuation,
is $\approx$5\% and is found for a hypothesised mass of $\approx$128~GeV.
The probability for such an excess to appear anywhere in the investigated mass range is around 40\%.

Exclusion limits on the inclusive production cross-section of a Standard Model
Higgs boson relative to the Standard Model cross-section are derived.
For this purpose, a modified frequentist approach $CL_s$~\cite{cls},
corresponding to the ratio of $p$-values for the signal-plus-background and
the background-only hypothesis for a given assumed signal
strength is used.
A given signal strength is excluded at 95\% confidence level if its $CL_s$  is smaller than 0.05.
The confidence levels are computed using a large number of signal-plus-background
pseudo-experiments, with different signal yields, and background-only pseudo-experiments.
The results, including systematic uncertainties, are shown in Figure~\ref{fig:Results}(b).
The impact of the experimental systematic uncertainties is about 5\% on the expected limit for the mass hypothesis of 120~GeV.

The expected median limit in the case of no signal varies from 3.3 to 5.8 as a function of the
Higgs boson mass.
The variations of the observed limit, between 2.0 and 5.8, are consistent with expected statistical fluctuations around the median
limit.

\section{Conclusions}

A search for the Standard Model Higgs boson in the  $H\rightarrow\gamma\gamma$ decay mode
has been performed using an integrated luminosity of 1.08~fb$^{-1}$ recorded by
the ATLAS experiment in 2011.
A high purity diphoton sample is selected.
No excess is found in the diphoton invariant
mass distribution in the mass range of 110-150~GeV.
The observed limit on the cross-section of a Standard Model-like Higgs boson decaying into a pair
of photons ranges between 2.0 and 5.8 times the Standard Model cross-section. These variations 
are compatible with statistical fluctuations around the expected limit for this data set.
These limits are the most stringent to date in this channel.

\section{Acknowledgments}

We thank CERN for the very successful operation of the LHC, as well as the
support staff from our institutions without whom ATLAS could not be
operated efficiently.

We acknowledge the support of ANPCyT, Argentina; YerPhI, Armenia; ARC,
Australia; BMWF, Austria; ANAS, Azerbaijan; SSTC, Belarus; CNPq and FAPESP,
Brazil; NSERC, NRC and CFI, Canada; CERN; CONICYT, Chile; CAS, MOST and
NSFC, China; COLCIENCIAS, Colombia; MSMT CR, MPO CR and VSC CR, Czech
Republic; DNRF, DNSRC and Lundbeck Foundation, Denmark; ARTEMIS, European
Union; IN2P3-CNRS, CEA-DSM/IRFU, France; GNAS, Georgia; BMBF, DFG, HGF, MPG
and AvH Foundation, Germany; GSRT, Greece; ISF, MINERVA, GIF, DIP and
Benoziyo Center, Israel; INFN, Italy; MEXT and JSPS, Japan; CNRST, Morocco;
FOM and NWO, Netherlands; RCN, Norway; MNiSW, Poland; GRICES and FCT,
Portugal; MERYS (MECTS), Romania; MES of Russia and ROSATOM, Russian
Federation; JINR; MSTD, Serbia; MSSR, Slovakia; ARRS and MVZT, Slovenia;
DST/NRF, South Africa; MICINN, Spain; SRC and Wallenberg Foundation,
Sweden; SER, SNSF and Cantons of Bern and Geneva, Switzerland; NSC, Taiwan;
TAEK, Turkey; STFC, the Royal Society and Leverhulme Trust, United Kingdom;
DOE and NSF, United States of America.

The crucial computing support from all WLCG partners is acknowledged
gratefully, in particular from CERN and the ATLAS Tier-1 facilities at
TRIUMF (Canada), NDGF (Denmark, Norway, Sweden), CC-IN2P3 (France),
KIT/GridKA (Germany), INFN-CNAF (Italy), NL-T1 (Netherlands), PIC (Spain),
ASGC (Taiwan), RAL (UK) and BNL (USA) and in the Tier-2 facilities
worldwide.





\bibliographystyle{elsarticle-num}
\bibliography{paper}

\begin{thebibliography}{10}
\expandafter\ifx\csname url\endcsname\relax
  \def\url#1{\texttt{#1}}\fi
\expandafter\ifx\csname urlprefix\endcsname\relax\def\urlprefix{URL }\fi
\expandafter\ifx\csname href\endcsname\relax
  \def\href#1#2{#2} \def\path#1{#1}\fi

\bibitem{higgs1}
{F.~Englert and R.~Brout}, Broken symmetry and the mass of gauge vector mesons,
  Phys.~Rev.~Lett 13 (1963) 321--323.

\bibitem{higgs2}
{P.~W.~Higgs}, Broken symmetries, massless particles and gauge fields,
  Phys.~Lett. 12 (1964) 132--133.

\bibitem{higgs3}
{G.~Guralnik, C.~Hagen and T.~Kibble}, Global conservation laws and massles
  particles, Phys.~Rev.~Lett. 13 (1964) 585--587.

\bibitem{lep}
{The LEP Working Group for Higgs boson searches, R.~Barate et al.}, {Search for
  the Standard Model Higgs boson at LEP}, Phys.~Lett. B565 (2003) 61--75.

\bibitem{tevatron1}
{The TEVNPH Working Group for the CDF and D0 Collaboration}, {Combined CDF and
  D0 upper limits on Standard Model Higgs boson production with up to
  6.7~fb$^{-1}$ of data}, {arXiv:1007.4587}.

\bibitem{tevatron2}
{The TEVNPH Working Group for the CDF and D0 Collaboration}, {Combined CDF and
  D0 upper limits on Standard Model Higgs boson production with up to
  8.6~fb$^{-1}$ of data}, {arXiv:1107.5518}.

\bibitem{atlas2010}
{ATLAS Collaboration}, {Limits on the production of the Standard Model Higgs
  boson in pp collisions at $\sqrt{s}$=7~TeV with the ATLAS detector}, accepted
  by Eur.~Phys.~J~C.

\bibitem{cms2010}
{CMS Collaboration}, {Measurement of the WW production and search for the Higgs
  boson in pp collisions at $\sqrt{s}$=7~TeV}, Phys.~Lett. B699 (2011) 25--47.

\bibitem{higgs-theory}
{LHC Higgs cross-section working group}, {Handbook of LHC Higgs cross-sections:
  1. Inclusive observables and references therein~}, {arXiv:1101.0593}.

\bibitem{tev-gg1}
{D0 Collaboration}, {Search for the standard model and a fermiophobic Higgs
  boson in diphoton final states}, {arXiv:1107.4587}.

\bibitem{tev-gg2}
{CDF Collaboration}, {Search for a Standard Model Higgs Boson Decaying Into
  Photons at CDF Using 7.0 fb$^{-1}$ of Data}, {CDF note
  CDF/PUB/EXOTIC/PUBLIC/10485}.

\bibitem{tev-gg3}
{The TEVNPH Working Group for the CDF and D0 Collaboration}, {Combined CDF and
  D0 Searches for the Standard Model Higgs Boson Decaying to Two Photons with
  up to 8.2 fb$^{-1}$}, {arXiv:1107.4960}.

\bibitem{ATLAS_detector}
{ATLAS Collaboration}, {The ATLAS experiment at the CERN Large Hadron
  Collider}, JINST 3 (2008) S08003.

\bibitem{ATLAS_EMC_calibration}
{ATLAS Collaboration}, {Expected Performance of the ATLAS Experiment, Detector,
  Trigger and Physics: Calibration and Performance of the Electromagnetic
  Calorimeter}, CERN-OPEN-2008-020 (2008) 44--71.

\bibitem{lumi2011}
{ATLAS Collaboration}, {Luminosity Determination in pp Collisions at
  $\sqrt{s}$=7~TeV using the ATLAS Detector in 2011}, ATLAS-CONF-2011-116.

\bibitem{powheg_gg}
S.~Alioli, P.~Nason, C.~Oleari, E.~Re, {NLO Higgs boson production via gluon
  fusion matched with shower in POWHEG}, JHEP (2009) 002.

\bibitem{mcnlo}
S.~Frixione, B.~R. Webber, {Matching NLO QCD computations and parton shower
  simulations}, JHEP 06 (2002) 029.

\bibitem{powheg_vbf}
P.~Nason, C.~Oleari, {NLO Higgs boson production via vector-boson fusion
  matched with shower in POWHEG}, JHEP 1002 (2010) 037.

\bibitem{pythia}
T.~Sjostrand, S.~Mrenna, P.~Z. Skands, {PYTHIA 6.4 Physics and Manual}, JHEP
  0605 (2006) 026.

\bibitem{Harlander:2002wh}
R.~V. Harlander, W.~B. Kilgore, {Next-to-next-to-leading order Higgs production
  at hadron colliders}, Phys. Rev. Lett. 88 (2002) 201801.

\bibitem{Anastasiou:2002yz}
C.~Anastasiou, K.~Melnikov, {Higgs boson production at hadron colliders in NNLO
  QCD}, Nucl. Phys. B646 (2002) 220--256.

\bibitem{Ravindran:2003um}
V.~Ravindran, J.~Smith, W.~L. van Neerven, {NNLO corrections to the total cross
  section for Higgs boson production in hadron hadron collisions}, Nucl. Phys.
  B665 (2003) 325--366.

\bibitem{Anastasiou:2008tj}
C.~Anastasiou, R.~Boughezal, F.~Petriello, {Mixed QCD-electroweak corrections
  to Higgs boson production in gluon fusion}, JHEP 04 (2009) 003.

\bibitem{deFlorian:2009hc}
D.~de~Florian, M.~Grazzini, {Higgs production through gluon fusion: updated
  cross sections at the Tevatron and the LHC}, Phys. Lett. B674 (2009)
  291--294.

\bibitem{Bolzoni:2010xr}
P.~Bolzoni, F.~Maltoni, S.-O. Moch, M.~Zaro, {Higgs production via vector-boson
  fusion at NNLO in QCD}, Phys. Rev. Lett. 105 (2010) 011801.

\bibitem{Brein:2003wg}
O.~Brein, A.~Djouadi, R.~Harlander, {NNLO QCD corrections to the
  Higgs-strahlung processes at hadron colliders}, Phys. Lett. B579 (2004)
  149--156.

\bibitem{Dawson:2002tg}
S.~Dawson, L.~H. Orr, L.~Reina, D.~Wackeroth, {Associated top quark Higgs boson
  production at the LHC}, Phys. Rev. D67 (2003) 071503.

\bibitem{Aglietti:2004nj}
U.~Aglietti, R.~Bonciani, G.~Degrassi, A.~Vicini, {Two-loop light fermion
  contribution to Higgs production and decays}, Phys. Lett. B595 (2004)
  432--441.

\bibitem{Actis:2008ug}
S.~Actis, G.~Passarino, C.~Sturm, S.~Uccirati, {NLO electroweak corrections to
  Higgs boson production at hadron colliders}, Phys. Lett. B670 (2008) 12--17.

\bibitem{Ciccolini:2007jr}
M.~Ciccolini, A.~Denner, S.~Dittmaier, {Strong and electroweak corrections to
  the production of Higgs + 2 jets via weak interactions at the LHC}, Phys.
  Rev. Lett. 99 (2007) 161803.

\bibitem{Ciccolini:2007ec}
M.~Ciccolini, A.~Denner, S.~Dittmaier, {Electroweak and QCD corrections to
  Higgs production via vector-boson fusion at the LHC}, Phys. Rev. D77 (2008)
  013002.

\bibitem{Ciccolini:2003jy}
M.~L. Ciccolini, S.~Dittmaier, M.~{Kr\"amer}, {Electroweak radiative
  corrections to associated WH and ZH production at hadron colliders}, Phys.
  Rev. D68 (2003) 073003.

\bibitem{Botje:2011sn}
M.~Botje, J.~Butterworth, A.~Cooper-Sarkar, A.~de~Roeck, J.~Feltesse, et~al.,
  {The PDF4LHC working group interim recommendations}, {arXiv:1101.0538}.

\bibitem{Lai:2010vv}
H.-L. Lai, M.~Guzzi, J.~Huston, Z.~Li, P.~M. Nadolsky, et~al., {New parton
  distributions for collider physics}, Phys.Rev. D82 (2010) 074024.

\bibitem{Martin:2009iq}
A.~D. Martin, W.~J. Stirling, R.~S. Thorne, G.~Watt, {Parton distributions for
  the LHC}, Eur. Phys. J. C63 (2009) 189--285.

\bibitem{Ball:2011mu}
R.~D. Ball, V.~Bertone, F.~Cerutti, L.~Del~Debbio, S.~Forte, et~al., {Impact of
  heavy quark masses on parton distributions and LHC phenomenology}, Nucl.Phys.
  B849 (2011) 296--363.

\bibitem{Djouadi:1997yw}
A.~Djouadi, J.~Kalinowski, M.~Spira, {HDECAY: A program for Higgs boson decays
  in the standard model and its supersymmetric extension}, Comput. Phys.
  Commun. 108 (1998) 56--74.

\bibitem{alpgen}
M.~L. Mangano, M.~Moretti, F.~Piccinini, R.~Pittau, A.~D. Polosa, {ALPGEN, a
  generator for hard multiparton processes in hadronic collisions}, JHEP 0307
  (2003) 001.

\bibitem{atlasSim}
{ATLAS Collaboration}, {The ATLAS simulation infrastructure}, Eur.~Phys.~J C70
  (2010) 823--874.

\bibitem{geant4}
{GEANT4 Collaboration, S.~Agostinelli et al.}, {GEANT4: A simulation toolkit},
  Nucl.~Instrum.~Meth.~A 506 (2003) 250--303.

\bibitem{egamma2010}
{ATLAS Collaboration}, {Electron performance measurements with the ATLAS
  detector using the 2010 LHC proton-proton collision data}, {to be submitted
  to Eur.~Phys.~J.~C}.

\bibitem{photon2010}
{ATLAS Collaboration}, {Measurement of the inclusive isolated prompt photon
  cross section in pp collisions at $\sqrt{s}$=7~TeV with the ATLAS detector},
  Phys.~Rev. D83 (2011) 052005.

\bibitem{Cacciari:area}
M.~Cacciari, G.~P. Salam, G.~Soyez, The catchment area of jets, JHEP 04 (2008)
  005.

\bibitem{crystalBall}
J.~E. Gaiser, {Charmonium spectroscopy from radiative decays of the J/$\Psi$
  and $\Psi^{\prime}$}, {Ph.D.~Thesis, SLAC-R-255 (1982)}.

\bibitem{diphoton2010}
{ATLAS Collaboration}, {Measurement of the isolated di-photon cross-section in
  pp collisions at $\sqrt{s}$=7~TeV with the ATLAS detector}, {submitted to
  Phys.~Rev.~D~}.

\bibitem{hqt}
{G.~Bozzi, S.~Catani, D.~de~Florian and M.~Grazzini}, {Transverse-momentum
  resummation and the spectrum of the Higgs boson at the LHC}, Nucl.~Phys. B737
  (2006) 73--120.

\bibitem{resbos1}
{C.~Balazs and C.-P.~Yuan}, {Higgs boson production at the LHC with soft gluon
  effects}, Phys.Lett. B478 (2000) 192--198.

\bibitem{resbos2}
{Q.-H.~Cao, C.-R.~Chen, C.~Schmidt and C.-P.~Yuan}, {Improved Predictions for
  Higgs $Q_T$ at the Tevatron and the LHC}, {arXiv:0909.2305}.

\bibitem{diphox}
{T.~Binoth, J.~Guillet, E.~Pilon, and M.~Werlen}, {A full next-to-leading order
  study of direct photon pair production in hadronic collisions}, Eur.~Phys.~J
  C16 (2000) 311.

\bibitem{resbos3}
{C.~Balazs, E.~L.~Berger, P.~M.~Nadolsky, and C.~P.~Yuan}, {Calculation of
  prompt diphoton production cross sections at Fermilab Tevatron and CERN LHC
  energies}, Phys. Rev. D76 (2007) 013009.

\bibitem{qmu}
{G.~Gowan, K.~Cranmer, E.~Gross and O.~Vitells}, {Asymptotic formulae for
  likelihood-based tests of new physics}, Eur.~Phys.~J C71 (2011) 1.

\bibitem{cls}
A.~L. Read, {Presentation of search results: The $CL_s$ technique}, J.~Phys.
  G28 (2002) 2693--2704.

\end{thebibliography}

\clearpage

\onecolumn
\begin{flushleft}
{\Large The ATLAS Collaboration}

\bigskip

G.~Aad$^{\rm 48}$,
B.~Abbott$^{\rm 111}$,
J.~Abdallah$^{\rm 11}$,
A.A.~Abdelalim$^{\rm 49}$,
A.~Abdesselam$^{\rm 118}$,
O.~Abdinov$^{\rm 10}$,
B.~Abi$^{\rm 112}$,
M.~Abolins$^{\rm 88}$,
H.~Abramowicz$^{\rm 153}$,
H.~Abreu$^{\rm 115}$
% ...
\bigskip

% ...
$^{10}$ Institute of Physics, Azerbaijan Academy of Sciences, Baku, Azerbaijan\\
$^{11}$ Institut de F\'isica d'Altes Energies and Departament de F\'isica de la Universitat Aut\`onoma  de Barcelona and ICREA, Barcelona, Spain\\
% ...
$^{48}$ Fakult\"{a}t f\"{u}r Mathematik und Physik, Albert-Ludwigs-Universit\"{a}t, Freiburg i.Br., Germany\\
$^{49}$ Section de Physique, Universit\'e de Gen\`eve, Geneva, Switzerland\\
% ...
$^{88}$ Department of Physics and Astronomy, Michigan State University, East Lansing MI, United States of America\\
% ...
$^{111}$ Homer L. Dodge Department of Physics and Astronomy, University of Oklahoma, Norman OK, United States of America\\
$^{112}$ Department of Physics, Oklahoma State University, Stillwater OK, United States of America\\
% ...
$^{115}$ LAL, Univ. Paris-Sud and CNRS/IN2P3, Orsay, France\\
% ...
$^{118}$ Department of Physics, Oxford University, Oxford, United Kingdom\\
% ...
$^{153}$ Raymond and Beverly Sackler School of Physics and Astronomy, Tel Aviv University, Tel Aviv, Israel\\
% ...
\end{flushleft}

\end{document}